\documentclass{amsart}
\usepackage{amsthm}
\usepackage[T1]{fontenc}
\usepackage[cp1250]{inputenc}
\usepackage{amssymb}
\usepackage{amsmath}
\usepackage{verbatim}
\usepackage{enumerate}
\newtheorem{thm}{Theorem}[section]

\newtheorem{prop}[thm]{Proposition}
\theoremstyle{definition}

\theoremstyle{remark}

\numberwithin{equation}{section}

\newcommand{\F}{\mathcal{F}}
\renewcommand{\t}{\tau}
\newcommand{\dx}{x-\tau(x)}
\newcommand{\dt}{\partial_\tau}
\newcommand{\D}{\mathcal{D}}
\newcommand{\Lam}{\mathbf{\Lambda}}

\newcommand{\A}{\mathbf{A}}
\newcommand{\prf}{{\bf Proof: }}
\newcommand{\norm}[1]{\left\Vert#1\right\Vert}
\renewcommand{\sc}[2]{\langle #1|#2 \rangle}
\newcommand{\R}{\mathbb R}
\newcommand{\No}{\mathbb{N}\cup\{0\}}
\newcommand{\Z}{\mathbb Z}
\newcommand{\C}{\mathbb C}
\newcommand{\N}{\mathbb N}
\renewcommand{\leq}{\leqslant}
\renewcommand{\geq}{\geqslant}
\renewcommand{\phi}{\varphi}
\newcommand{\dq}{\partial_q}
\newcommand{\be}{\begin{equation}}
\newcommand{\ee}{\end{equation}}
\newcommand{\bse}{\begin{subequations}}
\newcommand{\ese}{\end{subequations}}
\renewcommand{\H}{\mathcal{H}}
\newcommand{\tto}{\longrightarrow}
\renewcommand{\O}{\mathcal{O}}
\newcommand{\abs}[1]{\left\vert#1\right\vert}
\DeclareMathOperator{\id}{id}

\newcommand{\B}{\mathcal{B}}

\begin{document}

\begin{center}
{\Large\bf  Factorization method for second order functional equations
}

\bigskip

{\bf Tomasz Goliński\footnote{tomaszg@alpha.uwb.edu.pl},
Anatol Odzijewicz\footnote{aodzijew@labfiz.uwb.edu.pl}}

\end{center}
\begin{center} {Institute of Theoretical Physics\\
University in Bia{\l}ystok
\\Lipowa 41, 15-424 Bia{\l}ystok, Poland}
\end{center}
\bigskip

\begin{abstract}
We apply general difference calculus in order to obtain solutions to the functional
equations of the second order. We show that factorization method can be successfully
applied to the functional case. This method is equivariant under the change of variables.
Some examples of applications are presented.
\end{abstract}


\tableofcontents

\section*{Introduction}
In this paper we develop the theory of equations of the form
\be \label{0}\alpha(x)\psi(\t(x))+\beta(x)\psi(x)+\gamma(x)\psi(\t^{-1}(x))=\lambda\psi(x) \ee
generated by a bijection $\t:X\to X$ of the real line subset $X\subset \R$.

The equations of this type can be regarded as an alternate discretization of the second order differential
equations including Schr\"odinger
equation and generalization of difference and $q$-difference
equations. They also emerge from the change of variables in difference equations. On the other hand the functional
equations by themselves are of interest and have many important applications. They have been
investigated in many monographs and papers from the point of view of functional analysis, especially
$C^*$-algebras methods, see e.g. \cite{ant,arai}. We hope that our approach can be applied also in the numerical analysis of some
problems.

Our approach is based on $\t$-difference and $\t$-integral
calculus, see e.g. \cite{klimek}, which is a direct generalization of the standard one and is ideologically related
to the analysis on time scales. This gives
us a possibility to generalize the factorization method
elaborated effectively by many authors in the differential case, e.g. see \cite{darboux,infeld}.
Our approach allows one to consider this method in $q$-case, see \cite{Alanowa} for application,
and due to generality of $\t$ has the benefit that it is equivariant under the change of variables.

Section \ref{calculus} defines basic notions and operators of $\t$-calculus as well as formulates some relations between them.
In Section \ref{hilbert} we introduce and study Hilbert spaces with the scalar product given
by $\t$-integral on the orbits of the semigroup of natural numbers $(\No,+)$ or group of integer numbers $(\Z,+)$ acting on $X$
by the iterations of the bijection $\t$. Then, in Section \ref{factor} we construct the
chain of eigenproblems in these Hilbert spaces by the application of the factorization method. This construction gives a possibility to solve the
eigenproblems for the chain of Hamiltonians if one knows a solution to the initial one.

In Section \ref{cov} we discuss the change of variables and induced
transformation of Hilbert spaces, eigenproblems and parametrizing functions.
Some remarks about equivalence of one $\t$-difference operator to another are also presented.

In Section \ref{exampl} we show that our method includes the one investigated in the theory of
$q$-Hahn orthogonal polynomials, which are $q$-deformation of the classical ones, see \cite{OHT}.
In this section we also present some examples.

Some general facts about $\t$-equations is presented in Appendix A, i.e. some formulas for solutions. Reader can find there easy results which are the generalization of
theorems about $q$-Riccati equation from \cite{ala}.

Let us stress finally the importance of the functional equation \eqref{0} in the theory of discrete and $q$-discrete
polynomials, see \cite{OHT}. This link is also important to the integration of some quantum optical systems, see
\cite{HOT}.

\section{Calculus generated by the bijection of subset of the real numbers}\label{calculus}

Let us introduce preliminary definitions and facts of the calculus generated by the bijection
$\t:X\to X$ of the subset $X\subset \R$ (see also \cite{klimek}). Let $\F(X)$ denotes the algebra of all complex valued functions
$\phi:X\to\C$.
Let
$\mathcal O(x)=\{\t^n(x)|n\in\Z\}$
be the orbit of point $x$ under action of group $\Z$ given by $\t$ and $\O^+(x)=\{\t^n(x)|n\in\No\}$ --- orbit under action
of semigroup $\No$.

The subject of the paper is linked to the investigation of the following linear operators
\begin{equation} \label{T}T\phi(x):=\phi(\t(x)) \end{equation}
\begin{equation} \label{M}M_f \phi(x):=f(x)\phi(x) \end{equation}
\begin{equation} \label{dt}\dt\phi(x):=\frac{\phi(x)-\phi(\t(x))}{x-\t(x)}, \end{equation}
where $f\in\F(X)$.

The operator $\dt$ is a direct generalization of $q$-derivative $\dq$, which corresponds to the
special case when $\t(x)=qx$. We shall call it {\bf$\t$-derivative}. It satisfies the following analogue of Leibniz rule
\begin{equation} \dt(\phi\psi)=(T\phi)\dt\psi+\psi\dt\phi .\end{equation}
We define $\t$-integral as the solution of the equation
\begin{equation} \label{rown} \dt\phi=\psi \end{equation}
for given $\psi\in\F(X)$. Applying the operator $T^n$ to both sides of \eqref{rown} one obtains
\begin{equation}
\phi(\t^n(x))-\phi(\t^{n+1}(x))=(\t^n(x)-\t^{n+1}(x))\psi(\t^n(x)) \end{equation}
for $n=0,1,2,\ldots$, where $\t^n=\underbrace{\t\circ\ldots\circ\t}_{\textrm{n times}}$ and thus
\begin{equation}
\label{calk}\phi(x)-\phi(\t^\infty(x))=\sum_{n=0}^\infty (\t^n(x)-\t^{n+1}(x))\psi(\t^n(x)).\end{equation}
The formula
\eqref{calk} has sense only if $\t^n(x)$ has the limit when $n\to\infty$ and $\phi$ is continuous at
$\t^\infty(x)\in X$. Hence, if these conditions are satisfied, one can define the {\bf$\t$-integral}:
\begin{equation} \label{calkdef}\int_{\t^\infty(x)}^x \!\!\psi(t)d_\t t =\int_{\O^+(x)}\!\!\psi(t)d_\t t
:=\sum_{n=0}^\infty (\t^n(x)-\t^{n+1}(x))\psi(\t^n(x)). \end{equation}
Let $[a,b]\subset X$ be an interval such that $\t^\infty(a)=\t^\infty(b)$ then we shall define the integral of $\psi$ over $[a,b]$ by:
\begin{equation} \label{tint}\int_a^b \psi(t) d_\t t=\int_{\O^+(a)\cup\O^+(b)} \psi(t) d_\t t:= \int_{\t^\infty(b)}^b \!\!\psi(t) d_\t t\;-
\int_{\t^\infty(a)}^a \!\!\psi(t) d_\t t. \end{equation}
Let us note that integral is only formally over $[a,b]$ and it is based on $\O^+(a)\cup \O^+(b)$.
Similarly we will define $\t$-integral over $\O(x)$ by
\begin{equation} \label{tint_o}\int_{\O(x)}\!\!\psi(t) d_\t
t:=\sum_{n=-\infty}^\infty (\t^n(x)-\t^{n+1}(x))\psi(\t^n(x)).\end{equation}
The terminology introduced above is justified by the following properties of the $\t$-integral:
\begin{equation} \label{fund}\int_a^b (\dt \psi)(t) d_\t t=\psi(b)-\psi(a), \end{equation}
\begin{equation} \label{fund2}\dt \int_{\t^\infty(x)}^x \!\!\psi(t) d_\t t = \psi(x). \end{equation}
Moreover we have the following formula for the change of variables
\begin{equation} \label{chov}\int_a^b (T\psi)(t) \rho(t) d_\t t=\int_{\t(a)}^{\t(b)} \psi(t) (\dt \t^{-1})(t)(T^{-1}\rho)(t) d_\t t.\end{equation}

Let us denote by $\D(X)\subset\F(X)$ the
subalgebra of functions with finite support. This subalgebra is preserved by $T$ and $\dt$, as
well as by $M_f$.

For affine bijection $\t(x)=qx+h$ all formulae above reduce to the corresponding formulae of $q$-difference ($h=0$) or
$h$-difference ($q=1$) calculus. If $q\to 1$ ($h=0$) or $h\to 0$ ($q=1$) one obtains the formulae of the standard
differential and integral calculus. Formulae \eqref{fund} and \eqref{fund2} become then the fundamental theorem of
calculus and \eqref{chov} reduces to identity.

\section{Hilbert spaces with scalar product given by $\t$-integral}\label{hilbert}
Let us fix some weight function $\rho:X\to\R$. We can define the scalar product by
the $\t$-integral over the set $\mathcal S$
by the formula
\begin{equation} \label{sc}\sc{\psi}{\phi}:=\int_\mathcal S \overline \psi(x)\phi(x)\rho(x)d_\t x ,\end{equation}
where $\mathcal S$ is one of the following sets $\O(x_0), \O^+(a), \O^+(a)\cup\O^+(b)$ and
$\phi$,$\psi$  are complex valued functions defined on $X$.

For the positivity of \eqref{sc} we shall assume that $(\t^n(b)-\t^{n+1}(b))\rho(\t^n(b))\geq 0$ and
$(\t^n(a)-\t^{n+1}(a))\rho(\t^n(a))\leq 0$, $n\in\N\cup\{0\}$ for semigroup orbits case or
$(\t^n(x_0)-\t^{n+1}(x_0))\rho(\t^n(x_0))\geq 0$ for group orbit case.

Let us denote by $L^2(X,\rho d_\t)$
the Hilbert space of square-integrable functions in the sense of \eqref{sc}. It
is clear that $\D(X)\subset L^2(X,\rho d_\t)$ for any weight function $\rho$. One shall keep in mind that elements of this
space are classes of equivalence of functions from $\F(X)$, which can be identified with functions on $\mathcal S$.

If we consider semigroup orbit case $\O^+(a)\cup\O^+(b)$ and let $b=\t^N(a)$ for some $N\in\N$, then $L^2(X,\rho d_\t)$ is finite dimensional. This case
will not be discussed in the paper. So, it will be assumed everywhere that
$b\neq \t^N(a)$ for all $N\in\N$.

If one admits the case $\rho(x)=0$ in \eqref{sc} for some $x\in\mathcal S$, then the Hilbert space
reduces to direct sum of Hilbert spaces realized by the functions on orbits or finite sets. So, one comes back
to the cases defined above. The case of two semigroup orbits is also direct sum and it is considered here only due to the analogy with $q$-Hahn polynomials.

Let us now consider the operator $T$ acting in Hilbert space $L^2(X,\rho
d_\t)$ defined on $\D(X)$ by \eqref{T}. Note that $\D(X)$ is dense in $L^2(X,\rho d_\t)$.

\begin{prop}The adjoint operator $T^*$ is the weighted shift operator given by:
\bse\begin{equation} (T^*\phi)(x)=\left\{\begin{array}{ll} \mu(x)\phi(\t^{-1}(x)) & \textrm{if }x\neq a\textrm{ and }x\neq b\\
                            0 & \textrm{if }x=a\textrm{ or }x=b
\end{array}\right. \end{equation}
for the semigroup orbits case and
\begin{equation} (T^*\phi)(x)=\mu(x)\phi(\t^{-1}(x)), \end{equation} \ese
for the group orbit case, where $\phi\in \D_{T^*}$ and
\begin{equation} \label{mu}\mu(x):=\dt(\t^{-1})(x)\frac{\rho(\t^{-1}(x))}{\rho(x)}.\end{equation}
The domain $\D_{T^*}$ of $T^*$ contains $\D(X)$.
\end{prop}
Let us mention the following properties of the operators $T$ and $T^*$. For the group orbit case
one has identity
\bse\begin{equation} T^*T\phi=\mu\phi. \end{equation}
For the semigroup orbits case one obtains
\begin{equation} \label{TT} T^*T\phi=\mu {(1-\chi_{\{a,b\}})}\phi, \end{equation}
where $\chi_{\{a,b\}}$ is the characteristic function of the subset $\{a,b\}\subset X$ and
thus $M_{\chi_{\{a,b\}}}$ is a projector onto this subset.
For both group and semigroup orbits cases one has
\begin{equation} TT^*\phi=(\mu\circ \t)\phi .\end{equation} \ese

The operator $T$ is bounded if and only if $\sup\limits_{x\in\mathcal S} | \mu(x)|<\infty$ and then $\norm T=\norm{T^*}=(\sup\limits_{x\in\mathcal S} | \mu(x)|)^{\frac{1}{2}}$.

Let $\chi_a$ and
$\chi_b$ be the characteristic functions (indicators) of $\O^+(a)$ and $\O^+(b)$.  The orthogonal projectors
$M_{\chi_a}$ and $M_{\chi_b}$ map $\D(X)$ into itself. Since operator $T$ preserves subspaces
$L^2(\O^+(a),\rho d_\t)$ and $L^2(\O^+(b),\rho d_\t)$, it can be reduced to the components of the partition
\begin{equation} L^2(X,\rho d_\t)=L^2(\O^+(a),\rho d_\t)\oplus L^2(\O^+(b),\rho d_\t)\end{equation}
in the sense of \cite{akhiezer}. Thus we can consider each orbit separately.

Let us now consider the sequence of Hilbert spaces
$\H_k:=L^2(X,\rho_k d_\t)$ with the weight functions
$\rho_k$, $k\in\No$ related by the recurrences
\begin{equation} \label{rho_eta}\rho_{k+1}=\eta_k\rho_k \end{equation}
and
\begin{equation} \label{rho_B}\rho_{k+1}=T(B_k\rho_k) \end{equation}
for $B_k,\eta_k\in\mathcal F(X)$.
Since we demand two distinct formulae for $\rho_{k+1}$, we need to
add the consistency condition on $\eta_k$ and $B_k$, namely
\begin{equation} \label{pearson}T(B_k\rho_k)=\eta_k\rho_k \end{equation}
and in semigroup case we additionally impose a boundary conditions
\begin{equation} \label{bound}B_k(a)\rho_k(a)=B_k(b)\rho_k(b)=0 \end{equation}
for any $k\in\No$.

Let us denote by $g_k$ the ratio of $B_k$ and $B_{k+1}$:
\begin{equation} \label{B}B_{k+1}=g_k B_k .\end{equation}
Then we can easily see that formulas \eqref{rho_eta}--\eqref{bound} are automatically satisfied if
\eqref{pearson}--\eqref{bound} are valid for $k=0$ and
\begin{equation} \label{eta}\eta_{k+1}=T(g_k\eta_k) \end{equation}
for any $k\in\No$.

By introducing the function
\begin{equation} \label{A_def}A_k(x)=\frac{B_k(x)-\eta_k(x)}{x-\t(x)}\end{equation}
and expressing $\eta_k$ by $A_k$ and $B_k$, one can rewrite equation \eqref{pearson} in the form
\begin{equation} \label{tpearson}\dt(B_k\rho_k)=A_k\rho_k \end{equation}
In the case of $\t(x)=qx$ and when $q\to 1$, the equation \eqref{tpearson} corresponds to the well known Pearson
equation of the theory of classical orthogonal polynomials. This motivates us to call the consistency equation
\eqref{pearson} the \mbox{\bf$\t$-Pearson} equation.

We shall now consider the operators defined formally by \mbox{\eqref{T}-\eqref{dt}} as operators in Hilbert spaces. For $f\in\F(X)$ we
shall consider the operators of multiplication by the function $f$ as
\begin{equation} M_{f}:\H_k\to\H_{k+1}.\end{equation}
The operator $T$ acts in each of the spaces separately
\begin{equation} T:\H_k\to\H_k\end{equation}
and the $\t$-derivative becomes now the operator
\begin{equation} \label{op_dt}\dt=M_{(id-\t)^{-1}}(1-T):\H_k\to\H_{k+1}.\end{equation}
Since operators $M_{f}$, $T$ and $\dt$ are in general unbounded, we take $\D(X)$ as their domain and then
close them. We will denote their closure by the same symbols.
We also omit indices $k$ in symbols of operators $M$, $T$ and $\dt$ to simplify the notation.

In these settings we can apply equation \eqref{pearson} to calculate explicitly the term $\mu_k$ appearing in formula
\eqref{mu} for $T^*$:
\begin{equation} \label{mu2}\mu_k(x)=\dt(\t^{-1})(x)\frac{B_k(x)}{\eta_k(\t^{-1}(x))}.\end{equation}

\begin{prop}$\;$\\
The adjoint operators $M_f^*:\H_{k+1}\to \H_k$ and $\dt^*:\H_{k+1}\to\H_k$,
are given by
\begin{equation} \label{op_M*}M_{f}^*=M_{\eta_k\overline{f}} \end{equation}
\begin{equation} \dt^*=(1-T^*)M_{\eta_k(id-\t)^{-1}}. \end{equation}
The domains of $M_{f}^*$ and $\dt^*$ contain $\D(X)$.
\end{prop}
\prf By means of \eqref{rho_eta} and \eqref{sc} we get
$$\sc{f\phi}{\psi}_{k+1}=\int_\mathcal S \overline{f(x)\phi(x)}\psi(x)\rho_{k+1}(x)d_\t x=$$
$$=\int_\mathcal S \overline{\phi(x)}\psi(x)\overline{f(x)}\eta_k(x)\rho_{k}(x)d_\t x=
\sc{\phi}{\overline{f}\eta_k\psi}_{k}$$
for $\phi,\psi\in\D(\mathcal S)$. Formula for $\dt^*$ follows directly from \eqref{op_M*} and definition of $\dt$
\eqref{op_dt}.\qed

\section{Factorization method}\label{factor}

The proposed method is the direct generalization of method for the second
order differential equations, see \cite{darboux,infeld}. Instead of
the equation of the form
\begin{equation} \label{eq_func}\alpha(x)\psi(\t(x))+\beta(x)\psi(x)+\gamma(x)\psi(\t^{-1}(x))=\lambda\psi(x), \end{equation}
where $\lambda\in\C$ and $\alpha,\beta,\gamma\in\mathcal F(X)$, we will
consider the sequence (chain) of the factorized equations
\begin{equation} \label{t-rown} (\A_k^*\A_k\psi_k)(x)=\lambda_k\psi_k(x) \end{equation}
on $\psi_k\in\H_k$ with the operators $\A_k:\H_k\to \H_{k+1}$ and their adjoints $\A_k^*:\H_{k+1}\to \H_{k}$ \mbox{given by}
\bse \label{operatory}\begin{equation} \label{A}\A_k:=M_{h_k}\dt + M_{f_k} \end{equation}
\begin{equation} \A_k^*=(1-T^*)M_{h_k\eta_k(id-\t)^{-1}}+M_{\eta_k f_k},\end{equation} \ese
where $f_k:X\to\R$, $h_k:X\to\R$ and $k\in \No$.
We assume that $\A_k$ is a closure of the operator defined by \eqref{A} on $\D(X)$ and $\A_k^*$ is automatically closed. Then
by von Neumann theorem (see \cite{akhiezer}) $\A_k^*\A_k$ is self-adjoint and thus closed.
The essence of the approach is to postulate the relation
\begin{equation} \label{comm} \A_k\A_k^*=d_k \A_{k+1}^*\A_{k+1} + c_k,\end{equation}
where $c_k$, $d_k\in \C$ and $k\in\No$. It follows from \eqref{comm} that given a solution $\psi_k\in\H_k$ of equation \eqref{t-rown},
the function
\begin{equation} \label{new_sol}\psi_{k+1}:=\A_k\psi_k \in\H_{k+1}\end{equation}
is a solution of \eqref{t-rown} for $k+1$ with $\lambda_{k+1}=\frac{\lambda_k-c_k}{d_k}$. So, starting from the
equation \eqref{t-rown} for $k=0$ and its solution $\psi_0$, one generates the family of equations \eqref{t-rown} with
solutions given by
\begin{equation} \psi_k=\prod_{i=0}^{k-1}\A_i\psi_0.\end{equation}

The operators $\A_k$ and $\A_k^*$ depend on the functions $f_k$, $h_k$ as well as on $\eta_k$ and $B_k$ related
by the recurrences \eqref{eta} and \eqref{B} respectively.

Let us introduce the function
\begin{equation} \label{phi_def}\phi_k(x):=f_k(x)+\frac{h_k(x)}{x-\t(x)}, \end{equation}
which we will use instead of $f_k$.

\begin{prop} Condition \eqref{comm} is equivalent to the transformation rule
\begin{equation} \label{phi}\phi_{k+1}h_{k+1}=\frac{h_k}{d_k} \;\;T\!\!\left(\frac{\phi_k}{g_k}\right) \end{equation}
and the system of equations
\begin{eqnarray} \label{eq}
\frac{d_k g_k(x)B_k(x)(h_{k+1}(\t^{-1}(x)))^2}{(\dx)(\t^{-1}(x)-x)}-(\phi_k(x))^2\eta_k(x)+c_k=\nonumber\\
\frac{1}{d_kg_k(\t(x))}\left(\frac{d_kg_k(\t(x))B_k(\t(x))(h_k(x))^2}{(\t(x)-\t^2(x))(\dx)}-(\phi_k(\t(x)))^2\eta_k(\t(x))\frac{(h_k(x))^2}{(h_{k+1}(x))^2}\right)\end{eqnarray}
on the functions $B_k$, $\eta_k$, $\phi_k$, $g_k$, $h_k$, $h_{k+1}$ and constants $c_k$, $d_k$, where $k\in\No$.
\end{prop}
\prf
We compute explicitly left and right hand sides of \eqref{comm}.
By substituting the \eqref{operatory}, \eqref{phi} to \eqref{comm} and applying \eqref{TT} one has
\begin{eqnarray}  \label{LHS}\A_k\A_k^*=\frac{-h_k(x)\phi_k(\t(x))\eta_k(\t(x))}{\dx}T+\frac{(h_k(x))^2B_k(\t(x))}{(\t(x)-\t^2(x))(\dx)}+\nonumber\\
+(\phi_k(x))^2\eta_k(x)-
\frac{\phi_k(x)B_k(x)h_k(\t^{-1}(x))}{\dx}T^{-1} \end{eqnarray}
and
\begin{eqnarray}  \label{RHS}
\A_{k+1}^*\A_{k+1}=\frac{-\phi_{k+1}(x)\eta_{k+1}(x)h_{k+1}(x)}{\dx}T+\frac{B_{k+1}(x)(h_{k+1}(\t^{-1}(x)))^2}{(\dx)(\t^{-1}(x)-x)}+\nonumber\\
+(\phi_{k+1}(x))^2\eta_{k+1}(x)-\frac{B_{k+1}(x)\phi_{k+1}(\t^{-1}(x))h_{k+1}(\t^{-1}(x))}{\dx}T^{-1}. \end{eqnarray}
In \eqref{RHS} we have removed the projector $1-\chi_{\{a,b\}}$ appearing in \eqref{TT} for the case of orbits of semigroups. It
was multiplied by a function $B_{k+1}$, which by \eqref{bound} is equal to zero at the points $a$ and $b$ anyway.

Substituting \eqref{LHS} and \eqref{RHS} into \eqref{comm} and comparing coefficients in front of the operators $T$,
$\id$ and $T^{-1}$, one obtains
\begin{equation} \label{1}h_k(x)\phi_k(\t(x)\eta_k(\t(x))=d_kh_{k+1}(x)\phi_{k+1}(x)\eta_{k+1}(x),\end{equation}
\begin{eqnarray}
\label{2}\frac{B_{k}(\t(x))(h_k(x))^2}{(\t(x)-\t^2(x))(\dx)}+(\phi_k(x))^2\eta_k(x)=\nonumber\\
=d_k\left(\frac{B_{k+1}(x)(h_{k+1}(\t^{-1}(x)))^2}{(\dx)(\t^{-1}(x)-x)}+(\phi_{k+1}(x))^2\eta_{k+1}(x)\right)+c_k\end{eqnarray}
and
\begin{equation} \label{3}\phi_k(x) B(x)h_{k}(\t^{-1}(x))=d_kB_{k+1}(x)\phi_{k+1}(\t^{-1}(x))h_{k+1}(\t^{-1}(x)).\end{equation}

Using the transformation rules \eqref{eta} and \eqref{B} one sees that conditions \eqref{1} and \eqref{3}
are equivalent. They give the transformation rule \eqref{phi}
\begin{equation} \phi_{k+1}h_{k+1}=h_kT\left(\frac{1}{d_kg_k}\phi_k\right) \end{equation}
and condition \eqref{2} can be written in the form \eqref{eq}. Domains of $\A_k\A_k^*$ and $\A_{k+1}^*\A_{k+1}$ coincide
due to closeness.
\qed 

System of equations \eqref{eq} can be regarded as an equation for gauge functions $g_k$ (defined by \eqref{eta}-\eqref{B}) or
transformation rule for $h_k$ given in non explicit way.
Check Example \ref{ex_xi} for explicit formulae for $g_k$ in the case $h_k\equiv 1$ and $c_k=0$ as well as convergence conditions.
In the case $\t(x)=qx$ and $g_k=const$ solutions to \eqref{eq} have been classified in \cite{Alanowa}.

If we have the solution of \eqref{eq} (both $h_{k+1}$ and $g_k$) one can express the functions $(B_{k+1}, \eta_{k+1}, f_{k+1})$
of the $(k+1)^{st}$ step in the terms of
the previous functions $(B_{k}, \eta_{k}, f_{k})$.

Functions $B_0$, $\eta_0$, $\phi_0$ and $\lambda_0$ are related to initial equation \eqref{eq_func}
by
\begin{equation} \label{alpha}\alpha(x)=-\frac{\phi_0(x)\eta_0(x)h_0(x)}{\dx}, \end{equation}
\begin{equation} \beta(x)=(\phi_0(x))^2\eta_0(x)+\frac{B_0(x)h_0(\t^{-1}(x))}{(\dx)(\t^{-1}(x)-x)}, \end{equation}
\begin{equation} \gamma(x)=-\frac{B_0(x)\phi_0(\t^{-1}(x))h_0(\t^{-1}(x))}{\dx}, \end{equation}
and
\begin{equation} \label{lambda}\lambda=\lambda_0. \end{equation}
The formulae inverse to \eqref{alpha}-\eqref{lambda} have the form
\begin{equation} B_0(x)=\frac{(\dx)(\t^{-1}(x)-x)}{(h_0(\t^{-1}(x)))^2}\left(\beta(x)+(\dx)\alpha(x)\frac{\phi_0(x)}{h_0(x)}\right), \end{equation}
\begin{equation} \eta_0(x)=-\frac{(\dx)\alpha(x)}{\phi_0(x)h_0(x)}, \end{equation}
where $\phi_0$ and $h_0$ have to be chosen in such way that their ratio satisfies the equation of the $\t$-Riccati form \eqref{rhom}:
\begin{equation} \label{phi1}
\frac{\phi_0(\t(x))}{h_0(\t(x))}=\frac{-\frac{\gamma(\t(x))}{\dx}-\frac{\phi_0(x)}{h_0(x)}\beta(\t(x))}{\frac{\phi_0(x)}{h_0(x)}\alpha(\t(x))(\t(x)-\t^2(x))}.
\ee

By this approach we have reduced the problem of construction of the chain of Hamiltonians to finding the solution of \eqref{eq}
and choice of initial conditions is ruled by \eqref{phi1}.

\section{Change of variables}
\label{cov}

We are going to consider an invertible change of variables
\be \label{kappa}\kappa:X\tto Y.\ee
It is convenient to assume that $\kappa$ is homeomorphism in order to preserve the limits.
Such transform allows us to modify the function $\t$ in considered equations. Let us note that in differential equation
change of variables keeps the differentiation operators and only changes the direction of derivation and multiplies
them by some function factor. In discrete case one obtains in fact different derivations.

Change of variables defines the map from $\F(Y)$ to $\F(X)$ by
\be K:\F(Y)\ni f\tto Kf=f\circ\kappa\in\F(X).\ee
We can associate to $T$ the operator
\be\tilde T=K^{-1}TK\ee
acting as the shift by
\be \label{t'}\tilde\t=\kappa\circ\t\circ\kappa^{-1}\ee
on $\F(Y)$. One calls functions $\tau$ and $\tilde\tau$ equivalent if they are related by \eqref{t'} for some $\kappa$.
Operator $M_f$, $f\in\F(X)$, transforms to
\be K^{-1}M_fK=M_{f\circ\kappa^{-1}}.\ee
Thus similarity transformation $K^{-1}\cdot K$ maps shift operators
to shift operators and multiplication operators to multiplication operators.
However it turns out that derivation is preserved by this transform only up to the function factor:
\be K^{-1}\dt K=M_{(\dt\kappa)\circ\kappa^{-1}}\partial_{\tilde\t}.\ee

For example let $X=[0,\infty)$, $\t(x)=qx$, $Y=\R$ for $0<q\neq 1$. Then $\kappa(x)=\ln(x)$ transforms $\t$ to
\be \tilde\t(y)=y+\ln q\ee
and
\be K\dq K^{-1}=M_\frac{-\ln q}{(1-q)e^y} \partial_{\tilde\t} . \ee

Thus we can map $q$-difference equation to $h$-difference equation although there may appear function factor changing the form of equation.

For given $(X, \t)$ and $(Y, \tilde\t)$ in general it is not possible to construct such $\kappa$ that \eqref{t'} would be valid.
It is the case for example for $X=Y=[0,1]$, $\t(x)=x^2$ and $\tilde\t(y)=\frac25 y^3-\frac 35y^2+\frac65y$. It follows from the
fact that $\kappa$ maps fixed points of $\t$ into fixed points of $\tilde\t$. Since $\t$ has two fixed points ($0$ and $1$) while
$\tilde\t$ has three ($0$, $\frac12$ and $1$), thus $\tilde\tau$ is not equivalent to $\tau$.
In some cases however one can restrict
$X$ and $Y$ in such way that the function would become equivalent but it may happen that $\kappa$ is not given in terms of elementary functions
thus rending further calculations extremely complicated.

Let us now consider change of variables defined by \eqref{kappa} from the point of view of Hilbert spaces and introduced
operators and functions. Let us consider Hilbert spaces $\tilde \H_k$ of functions on the set $Y$ defined in such way that
it makes $K$ an
unitary operator mapping $\H_k$ to $\tilde\H_k$. Simple calculations shows that we have to consider the spaces
$\tilde \H=L^2(\kappa(X),\tilde\rho_k d_{\tilde\t})$ with weight given by
\be \tilde \rho_k=(\partial_{\tilde\t}\kappa^{-1})\;\;K^{-1}\rho_k . \ee
In order to keep formulas for transformation of $\tilde \rho_k$ the same as \eqref{rho_eta} and \eqref{rho_B} one has to define
\be \tilde B_k =\frac{\tilde T^{-1}(\partial_{\tilde\t}\kappa^{-1})}{(\partial_{\tilde\t}\kappa^{-1})} K^{-1} B_k\ee
and
\be \tilde \eta_k= K^{-1}\eta_k.\ee
These functions automatically satisfy $\tilde\t$-Pearson equation and obey the same transformation laws with function
\be\tilde g_k=K^{-1}g_k.\ee
If we put the following transformation rules for functions $f_k$ and $h_k$
\be \label{tildeh}\tilde h_k = K^{-1}\left(h_k\dt(\kappa)\right)\ee
\be \tilde f_k = K^{-1}f_k,\ee
then operators $\tilde\A_k=K^{-1}\A_k K$ and $\tilde\A_k^*=K^{-1}\A_k^* K$ are of the same form as original with $\t$ replaced by $\tilde\t$.
Clearly $\tilde\A_k^*$  is an adjoint of $\tilde\A_k$ since $K$ is an unitary.
Let us note that due to \eqref{tildeh} the condition $h_k\equiv 1$ is not preserved by the change of variables except
in the case of translation $\kappa(x)=x+c$, $c\in\R$.

Solutions of considered second order equation \eqref{t-rown} transform as follows
\be \label{cov_sol}\tilde\psi_k=K^{-1}\psi_k.\ee

Thus we were able to map the chain of Hamiltonians on $X$ to the chain of Hamiltonians on $Y$ preserving all relations
introduced in Section \ref{hilbert} and \ref{factor}. This shows that considered method is equivariant with respect to
the change of variables. It was not true for the case of $q$-difference equations.

\section{Some examples}
\label{exampl}
We are going to present several examples for the application of factorization method described above.

\subsection{Particular solution of \eqref{eq}} \label{ex_xi}

We are going to construct a solution equation \eqref{eq} in terms of $\t$-integrals by application of propositions from
Appendix \ref{ap_ric}.

We consider the simplest case $h_k\equiv 1$. Let us introduce the function
\begin{equation} \label{xi_def}  \xi_k(x):=(\phi_k(x))^2\eta_k(x)-\frac{d_k g_k(x) B_k(x)}{(\dx)(\t^{-1}(x)-x)}. \end{equation}
After substituting $\xi_k$ instead of $g_k$ the equation \eqref{eq} takes the form
\begin{equation} \label{xi}\xi_k(x)=\frac{\left(\frac{
B_k(\t(x))}{(\t(x)-\t^2(x))(\dx)}-c_k\right)\xi_k(\t(x))+c_k(\phi_k(\t(x)))^2\eta_k(\t(x))}{(\phi_k(\t(x)))^2\eta_k(\t(x))-\xi_k(\t(x))},\end{equation}
which is also a $\t$-Riccati equation \eqref{rhom}. It is more convenient than the form \eqref{eq} because in
the case of $c_k=0$ one can apply the Proposition \ref{triangle} from Appendix \ref{ap_ric} which allows one to write down formulae for solutions.
To that end we have to invert the fractional map on right-hand side of \eqref{xi} and use it to define matrix
$\Lam_k(x)$ appearing in \eqref{r2}. We get:
\begin{equation} \Lam_k(x)=\left(\begin{array}{@{}cc@{}} 1 &
\frac{(x-\t(x))(\t(x)-\t^2(x))}{B_k(\t(x))} \\ 0 &
\frac{(x-\t(x))(\t(x)-\t^2(x))\phi_k(\t(x))^2\eta_k(\t(x))}{B_k(\t(x))} \end{array}\right).\end{equation} Since this
matrix can be singular at $\t^\infty(x)$, we perform $\t$-Darboux transform \eqref{dar_sing} to $\Lam_k(x)$ with
$\delta_2=\lambda$ and $\delta_1=0$. In order to apply Proposition \ref{converg} we have to compute $\tilde\Lam'_k$
defined by \eqref{lamtylda}:
\begin{equation} \tilde\Lam'_k(x)=\left(\begin{array}{@{}cc@{}} 0 & -
\frac{(\t(x)-\t^2(x))(x-\t^\infty(x))^\lambda}{B_k(\t(x))} \\
0 &
\frac{1}{x-\t(x)}-\frac{\phi_k(\t(x))^2\eta_k(\t(x))}{B_k(\t(x))}\frac{(\t(x)-\t^2(x))(x-\t^\infty(x))^\lambda}{(\t(x)-\t^\infty(x))^\lambda}
\end{array}\right).\end{equation}
Let us put $\lambda=-1$. If the functions $f_k$, $\eta_k$ and $B_k$ are regular and non-vanishing at
$\t^\infty(x)$ and $\t$ is differentiable at $\t^\infty(x)$ then we have
\begin{equation}\lim_{n\to\infty}\tilde\Lam_k'(\t^n(x))=\left(\begin{array}{@{}cc@{}} 0 &
-\left(\frac{d\t}{dx}(\t^\infty(x))-\frac{d\t^2}{dx}(\t^\infty(x))\right)(B_k(\t^\infty(x)))^{-1}\\
0 & 0 \end{array}\right).\end{equation}
In such a  way the singularity in $\tilde\Lam_k$ is removed. If
we assume that $\t$ satisfies the condition $i)$ of Proposition \ref{converg} then
the resolvent matrix is well defined and by Proposition
\ref{triangle} we find
\begin{equation} {\Lam'_k}_\infty(x)=\left(\begin{array}{@{}cc@{}} 1 & F_k(x) \\
                                            0 & \exp \left(\int^x_{\t^\infty(x)}\frac{\ln
\frac{(\t(t)-\t^\infty(t))(t-\t(t))(\t(t)-\t^2(t))\phi_k(\t(t))^2\eta_k(\t(t))}{(t-\t^\infty(t))B_k(\t(t))}
                                            }{t-\t(t)}d_\t t\right) \end{array}\right),\end{equation}
where $F_k(x)$ is given by
\begin{eqnarray}  F_k(x)=\exp \left(\int^x_{\t^\infty(x)}\!\!\frac{\ln
\frac{(\t(t)-\t^\infty(t))(t-\t(t))(\t(t)-\t^2(t))\phi_k(\t(t))^2\eta_k(\t(t))}{(t-\t^\infty(t))B_k(\t(t))} 
}{t-\t(t)}d_\t t \right)\times\nonumber\\
\times\!\int^x_{\t^\infty(x)}\!\!\frac{
(\t(t)-\t^2(t))}{(t-\t^\infty(t))B_k(\t(t))                             
}\times\nonumber\\
\!\!\!\!\times\exp \!\!\left(\int^t_{\t^\infty(x)}\!\!\!\!\frac{\ln
\frac{B_k(\t(s))}{(\t(s)-\t^\infty(s))(s-\t^\infty(s))(s-\t(s))(\t(s)-\t^2(s))\phi_k(\t(s))^2\eta_k(\t(s))} 
}{s-\t(s)}d_\t s \!\right)d_\t t .\end{eqnarray}
Thus, by \eqref{sol} and \eqref{meroF} one gets
\begin{eqnarray}\label{xic0}
\xi_k(x)=(x-\t^\infty(x))^{-1}\exp \left(-\int^x_{\t^\infty(x)}\!\!\frac{\ln
\frac{(\t(t)-\t^\infty(t))(t-\t(t))(\t(t)-\t^2(t))\phi_k(\t(t))^2\eta_k(\t(t))}{(t-\t^\infty(t))B_k(\t(t))} 
}{t-\t(t)}d_\t t \right)\times\nonumber \\
\times\frac{1}{{\xi_k}_0^{-1}\!-\!\!
\int^x_{\t^\infty(x)}\!\frac{
(\t(t)-\t^2(t))}{(t-\t^\infty(t))B_k(\t(t))                             
}
\exp\!\!\left(\!\!\int^t_{\t^\infty(x)}\!\!\!\!\frac{\ln
\frac{B_k(\t(s))}{(\t(s)-\t^\infty(s))(s-\t^\infty(s))(s-\t(s))(\t(s)-\t^2(s))\phi_k(\t(s))^2\eta_k(\t(s))} 
}{s-\t(s)}d_\t s \!\right)\!\!d_\t t }.\end{eqnarray}

From \eqref{xic0} by application of \eqref{xi_def}, \eqref{phi_def}, \eqref{phi}, \eqref{B},\eqref{eta} and \eqref{operatory}
we can write down explicit form of operators $\A_{k+1}$ and $\A_{k+1}^*$. We will not do it here due to length of formulas.


\subsection{$q$-Hahn polynomials}

Let us show now the well-know application of the method described in a previous section to $q$-deformed classical polynomials.
We consider $\t(x)=qx$ for $0<q<1$ and to stress it we replace the symbol $\dt$ and $T$ by $\dq$ and $Q$ respectively.
We assume that $B_0$ and $A_0$ defined by \eqref{A_def} are the second and first order polynomials respectively. We
consider a function $\rho_0$ satisfying $q$-Pearson equation \eqref{tpearson} and we put $f_0\equiv0$, $h_0\equiv1$. In this case the
equation \eqref{eq_func} is known as Hahn equation. Its solutions are given by basic hypergeometric series, see
\cite{hahn,rahman}, and we know from the theory of $q$-classical orthogonal polynomials that for certain choices of
$\lambda_0$, we obtain polynomials which belong to $\H_0$ and form orthogonal system with respect to weight
function $\rho_0$, see \cite{OHT}.

We want to keep the same type of equation for all $k$, i.e. we require that all $f_k\equiv0$, $h_k\equiv1$. We shall find the conditions
under which this requirement can be satisfied along with relation \eqref{comm}.

Equation \eqref{phi} reduces itself to
\begin{equation} g_k d_k=q^{-1}. \end{equation}
The transformation rule for $\eta_k$ \eqref{eta} can be rewritten in terms of $A_k$ as follows
\begin{equation} A_{k+1}(x)=q Q A_k(x)+\dq B_{k+1}(x). \end{equation}
Thus $A_k$ and $B_k$ remain a polynomials of degree one or two respectively for all $k$.

We can express function $\xi_k$ defined by \eqref{xi_def} in terms of our variables as
\begin{equation} \xi_k=\frac{A_k}{(1-q)x} \end{equation}
and equation \eqref{xi} is satisfied for $c_k=-\dq A_k$, which is a constant for all $k$. Let us mention that since
$\xi_k$ is singular we can obtain this solution from \eqref{xi} by the product \eqref{sol} described in Appendix \ref{ap_ric} only by applying singular
Darboux transform \eqref{dar_sing} with $\delta_1-\delta_2=1$.

Formula \eqref{new_sol} yields that if $\{P^k_n\}$ is an OPS for $\rho_k$ then $\{\dq P^k_n\}$ is an OPS for
$\rho_{k+1}$. We can reverse this procedure. Namely, by applying $\A_k^*$ to both sides of \eqref{new_sol} we gather
that
\begin{equation} \psi_k=\frac{1}{\lambda_k}\A_k^*\psi_{k+1}. \end{equation}
This allows us to express any polynomial by the formula
\begin{equation} P^k_n=\mu^k_n \A_{k+1}^*\ldots \A_{k+n}^* 1, \end{equation}
where $\mu^k_n\neq 0$ are  some normalization constants, see \cite{OHT}.

We can generalize this result by dropping the requirement that $f_k\equiv 0$ for $k\geq 1$ and then put $c_k=0$.
Then we can solve the equation \eqref{xi}
by the formula \eqref{xic0}. We can express that solution by basic hypergeometric function. This case is presented as
an example in \cite{GO}.

By the application of the change of variables described in Section \ref{cov} one can obtain solutions to another family of
equations with $\t(x)\neq qx$. For example, let $\kappa(x)=e^x$. Then $\tilde\tau(y)=y^q$. Hahn equation transforms into
\be \tilde \A^*_k \tilde\A_k \psi_k=0, \ee
where
\be \A^*_k \A_k =\left(\frac{1-e^q}{1-q}\frac{y}{\ln y}K^{-1}A_k + \left(\frac{1-e^q}{1-q}\right)^2 \left(\partial_{\tilde\tau}
\frac{qy^{\frac{1}{q}}}{\ln y} \right)K^{-1}B_k \right)+\ee
$$+\left(\frac{1-e^q}{1-q}\right)^2 \frac{y}{\ln y} K^{-1}B_k \partial_{\tilde\tau}\tilde T^{-1} \partial_{\tilde\tau}.$$
Its solutions are related to orthogonal polynomials by formula \eqref{cov_sol} and are orthogonal with respect to the
scalar product with weight function
\be \tilde\rho_k(y)=\frac{\ln y}{y-y^q}(1-q) K^{-1}\rho_k(y).\ee

\subsection{The case $g_k=\rm{const}$ and $\t(x)=qx$}

This example is based on the paper \cite{Alanowa}.
We assume that
\begin{equation} \t(x)=qx\quad\textrm{for}\quad 0<q<1, \end{equation}
\be h_k\equiv 1,\ee
\begin{equation} g_k=q^{-2},\end{equation}
\begin{equation} d_k=1 \end{equation}
and that $B_0$ is a constant function. From \eqref{B} we gather that
\begin{equation} B_k=q^{-2k}B_0. \end{equation}
Let $\alpha_k:=(\phi_k)^2\eta_k$. Equation \eqref{eq} now takes the form
\begin{equation}\label{ex2_eq}
\alpha_k(x)=q^{-2}\alpha_k(qx)-c_k. \end{equation} By substituting into equation \eqref{ex2_eq} the series
\begin{equation} \alpha_k(x)=x^\lambda\sum_{n\in\Z} a_n x^n,\end{equation}
where $\lambda\in[0,1)$, we obtain
\begin{equation} \alpha_k(x)=b_kx^{-2}-\frac{c_k}{1-q^2} \end{equation}
for any $b_k\in\C$. In order to preserve the transformation rules \eqref{eta} and \eqref{phi} we have to put
$b_k=q^{4k}b_0$ and $c_k=q^{2k}c_0$. Under these assumptions the operators $\A_k$ and $\A_k^*$ take the form
\begin{equation} \A_k=-\frac{1}{(1-q)x}Q+q^{2k}(Q^{k}\phi_0), \end{equation}
\begin{equation} \A_k^*=\frac{q^{2k}b_0x^{-2}-\frac{c_0}{1-q^2}}{(Q^k\phi_0)}-\frac{B_0}{q^{2k}(1-q)x}Q^{-1}. \end{equation}
Thus we have solved the factorization problem for the operators
\begin{eqnarray} \A_k^*\A_k=\left(x^{-1}\frac{q^{2k}c_0}{\phi_k(x)(1-q)^2(1+q)}- x^{-3}\frac{q^{4k}b_0}{\phi_k(x)(1-q)}\right)
Q-\nonumber \\
-x^{-1}\frac{B_0\phi_k(q^{-1}x)}{q^2(1-q)}Q^{-1}+x^{-2}\left(q^{4k}b_0+\frac{
B_0}{q(1-q)^2}\right)-\frac{q^{2k}c_0}{(1-q^2)}.\end{eqnarray}
The relation
\begin{equation} \label{kerA*} \A_{k-1}^*\psi_k=0 \end{equation}
implies that $\A_{k}^*\A_{k}\psi_{k}=(\A_{k-1}\A_{k-1}^*-c_{k-1})\psi_{k}=-c_{k-1}\psi_k$. Thus $\psi_k$ is an
eigenvector of $\A_k\A_k^*$ with the eigenvalue $-c_{k-1}$. We easily find the solution to \eqref{kerA*} by the use of
\eqref{prodcalk}:
\begin{equation} \label{ex-q-calk}\psi_k=\kappa\exp{\int_0^x\frac{\ln\left(\frac{(1-q)qt\alpha_k(t)}{B_k\phi_k(qt)}\right)}{t(1-q)}d_qt},\end{equation}
where  $\kappa$ is a constant. Let us assume that \eqref{ex-q-calk} is convergent.
We have to check if $\psi_k\in\H_k$, i.e. if it is square integrable with weight
function $\rho_k$ emerging from $q$-Pearson equation \eqref{pearson}. To that end we shall calculate function
$\psi_k^2\rho_k$. Let us rewrite Pearson equation in the form
\begin{equation} \frac{\alpha_k}{\phi_k^2B_k}=\frac{Q\rho_k}{\rho_k}.\end{equation}
From \eqref{kerA*} we conclude that
\begin{equation} \frac{Q\psi_k}{\psi_k}=\frac{B_k}{(1-q)q^3x}\frac{\alpha_k}{\phi_k}\end{equation}
and thus
\begin{equation} \psi_k^2\rho_k=\left(b_k-\frac{c_k}{1-q^2}x^2\right)\frac{(1-q)^2q^6}{B_k} \quad Q(\psi_k^2\rho_k).\end{equation}
We can rewrite this equation in the following form
\begin{equation} \psi_k^2\rho_k=\gamma\left(1-\frac{x}{\beta}\right)\left(1+\frac{x}{\beta}\right)\quad Q(\psi_k^2\rho_k),\end{equation}
where $\beta=\sqrt\frac{b_k(1-q^2)}{c_k}$ and $\gamma=-\frac{(1-q)^2q^6b_k}{B_k}$. If we request that $\gamma=1$
by appropriate choice of $b_k$ and $B_k$
then we have the following solution
\begin{equation} (\psi_k(x))^2\rho_k(x)=\mu\left(-\frac{x}{\beta};q\right)_\infty\left(\frac{x}{\beta};q\right)_\infty, \end{equation}
where
\begin{equation} \label{qfact}(\alpha;q)_n:=(1-\alpha)(1-q\alpha)\cdots(1-q^{n-1}\alpha) \end{equation}
and $\mu\in\R\setminus\{0\}$ is such that the scalar product is positively definite.
Let us estimate the $q$-integral of this function.
\begin{equation} \int^a_0 (\psi_k(x))^2\rho_k(x)d_qx=\mu(1-q)a\sum_{n=0}^\infty q^n \left(-\frac{q^na}{\beta};q\right)_\infty\left(\frac{q^na}{\beta};q\right)_\infty\end{equation}
Since $(\pm a/\beta;q)_n$ is convergent when $n\to\infty$ for $0<q<1$ then
$(\pm aq^n/\beta;q)_\infty$ is necessarily convergent to $1$. Thus
it is bounded from below and above. Let us denote these lower bounds by $m_\pm$ and upper bounds by $M_\pm$. Thus
$$ \abs{\int^a_0 (\psi_k(x))^2\rho_k(x)d_qx}\leq \abs{\mu a} \max(\abs{M_-},\abs{m_-})\max(\abs{M_+},\abs{m_+})(1-q)\sum_{n=0}^\infty
q^n=$$
\be=\abs{\mu a} \max(\abs{M_-},\abs{m_-})\max(\abs{M_+},\abs{m_+}).\ee
So this integral is finite as well as
$\sc{\psi_k}{\psi_k}_k=\int_a^b \psi_k^2\rho_kd_qx$ and thus $\psi_k\in\H_k$ for any choice of limits of integration
fulfilling \eqref{bound} and constants such that $\gamma=1$.

Thus by \eqref{new_sol} we have
\begin{equation} \psi_k^N=\A_{k+N-1}\cdot\ldots\cdot \A_{k+1}\A_k\psi_k \in\H_{k+N}\end{equation}
and $\psi_k^N$ is an eigenvector of the operator
$\A_{k+N}^*\A_{k+N}$
with the eigenvalue equal to
\begin{equation} \lambda_k^N=-\!\!\!\sum_{l=k-1}^{k+N-1} c_l.\end{equation}
For example, if we put
\begin{equation} \phi_0(x)=\frac{b_0(1-q)q^2}{B_0x e^{q^{-s}x^s}}-\frac{c_0(1-q)x}{B_0e^{q^{-s}x^s}}, \end{equation}
then
\begin{equation} \psi_0(x)=\frac{\kappa(1-q^s)}{(1-q)^2}x^s. \end{equation}

Examples given above show the importance of the factorization method in the theory of the
orthogonal polynomials and quantum mechanical spectral problems. In \cite{Alanowa}
the wide class of $q$-Schr\"odinger equations, which are solved by this method, is presented.

\subsection{The case $g_k=\rm{const}$ and $\t$ is a fractional map}

This example is similar to previous one but $\t$ is kept arbitrary. We still consider the case
when $h_k\equiv 1$ and $g_k=\operatorname{const}$, $k\in\No$. From equation \eqref{eq}
one has that that case is possible for example if
\be B_k(x)=b_k (x-\t(x))(\t^{-1}(x)-x)\ee
and
\be (\phi_k(x))^2\eta_k(x)=a_k\ee
for some constants $a_k$, $b_k$. These conditions are compatible with transformation rules \eqref{eta},\eqref{B} and \eqref{phi} by
following relations
\be a_{k+1}=\frac{g_k}{d_k^2} a_k \ee
\be b_{k+1}=g_k b_k.\ee
The equations \eqref{eq} reduce to quadratic equations with constant coefficient and constants $g_k$ are given as its roots.
We will treat function $\phi_0$ as arbitrary as well as constants $a_0$, $b_0$, $c_k$, $d_k$.
In this case the equations \eqref{t-rown} assume the form
\be  \label{ex_eq}\frac{-a_k}{\phi_k(x)(x-\t(x))}\psi_k(\t(x))+(b_k+a_k-\lambda_k)\psi_k(x)- b_k\phi_k(\t^{-1}(x))(\t^{-1}(x)-x)\psi_k(\t^{-1}(x))=0 .\ee
Let us consider the equations
\be \label{kerA}\A_k\psi_k=0.\ee
It takes the form
\be \psi_k(x)=\frac{1}{(x-\t(x))\phi_k(x)}\psi_k(\t(x))\ee
and
\be \label{alph_eq}(\psi_k(x))^2\rho_k(x)=\frac{b_k}{a_k}\dt\t(x) \;\;\;(\psi_k(\t(x)))^2\rho_k(\t(x)).\ee
If we postulate the following form of the solution to \eqref{alph_eq}
\be \label{alpha_sing}(\psi_k(x))^2\rho_k(x)=(x-\t(x))^\lambda \alpha_k(x),\ee
where $\lambda\in \Z$, then we get
\be \label{r_alpha}\alpha_k(x)=\frac{b_k}{a_k}(\dt\t(x))^{\lambda+1} \alpha_k(\t(x)).\ee
Necessary condition for this equation to have non-zero solution is
\be \left(\frac{b_k}{a_k}\right)^{\frac1{\lambda+1}}=\dt\t(\t^\infty(x)).\ee

Let us now consider as an example the fractional map
\be \label{homog} \t(x)=\frac{ax}{(a-1)x+1}\ee
for $a>0$, $a\neq 1$, which preserves the interval $(0,1)$ and has two fixed points --- $0$ and $1$.
Let us set $\mathcal S=\O(\frac12)$.
One can find formula for $\t^k$:
\be \t^k(x)=\frac{a^kx}{(a^k-1)x+1}. \ee
It is easily computed that
\be (\dt \t)(x)=\frac{a}{(-1+a^2)x+1}\ee
and
\be (\dt \t)(\t^k(x))=a \frac{(a^k-1)x+1}{(a^{k+2}-1)x+1}.\ee
Iteration of \eqref{r_alpha} gives us that
\be \alpha_k(x)=\left\{\begin{array}{ll} \left(\frac{(a-1)x+1}{(-x+1)^2}\right)^{\lambda+1} \alpha_k(0) & 0<a< 1 \\
\left(\frac{(a-1)x+1}{ax^2}\right)^{\lambda+1} \alpha_k(1) & a>1\end{array}\right. .\ee
For $\lambda\neq-1$ the function $(\psi_k(x))^2\rho_k(x)(x-\t(x))$ has the pole at $0$ or $1$, while
 for $\lambda=-1$ it is constant. Thus $\psi_k\notin\H_k$ for any $k\in\No$. It is also possible to show that
 kernel of $A^*_{k+1}$ also is empty.

We can still consider solutions to \eqref{kerA} as functions and use \eqref{new_sol}
to construct new solutions to \eqref{ex_eq}.
For example, let us fix
\be \phi_0(x)=\frac{1}{x-\t(x)}.\ee
From \eqref{phi} and assumptions of this example it follows that
\be \phi_k(x)=\frac{1}{D_k G_k} \frac{1}{\t^k(x)-\t^{k+1}(x)},\ee
where $G_{k}=g_{k}\cdots g_{0}$ and $D_{k}=d_{k}\cdots d_{0}$. Equations \eqref{kerA} assume the form
\be \psi_k(x)=D_kG_k\frac{\t^k(x)-\t^{k+1}(x)}{x-\t(x)}.\ee
Its solutions can be found by iteration and are given by
\be \psi_k(x)=\prod_{j=0}^{k-1} \frac{((a^j-1)x+1)((a^{j+1}-1)x+1)}{(x-1)^2}\psi_k(0)\ee
for $0<a<1$ and $(G_kD_k)^\frac{1}{k}a=1$ or
\be \psi_k(x)=\prod_{j=0}^{k-1} \frac{((a^j-1)x+1)((a^{j+1}-1)x+1)}{a^{2j+1}x^2}\psi_k(1)\ee
for $a>1$ and $(G_kD_k)^\frac{1}{k}=a$.

From these solutions we can construct new by \eqref{new_sol}.



\appendix
\section{$\t$-Riccati equation and $\t$-Darboux transform}
\label{ap_ric}

We will present here additional information about $\t$-difference equations and methods of solving.
Most of this appendix is a direct generalization of \cite{ala}.

Let us begin this section by solving the equation
\begin{equation} \dt \phi=\phi\end{equation}
with initial condition $\phi(\t^\infty(x))=1$, under he assumption that $\t$ has the contracting property $|\t(x)-\t(y)|\leq M |x-y|$, where $0<M<1$.
Then, by the iteration one finds
\begin{equation} \label{exp}\phi(x)=\exp_\t(x):=\prod_{n=0}^\infty \frac{1}{1-\t^n(x)+\t^{n+1}(x)}. \end{equation}
We shall call $\exp_\t$ the \mbox{\bf$\t$-exponential} function.

Let us formulate one more identity which was used throughout the paper.
\begin{equation} \label{prodcalk}\prod_{n=0}^\infty \bigg(T^n F(x)\bigg) = \exp\left( \int^x_{\t^\infty(x)}\!\frac{\ln F(t)}{t-\t(t)}d_\t t\right). \end{equation}
It follows from the observation that
\be \ln \prod_{n=0}^\infty \bigg(T^n F(x)\bigg) = \sum_{n=0}^\infty T^n \ln F(x)\ee
and from the definition of $\t$-integral \eqref{tint}.
Let us note that $\exp$ and $\ln$ appearing in \eqref{prodcalk} are standard exponent and logarithm.

From \eqref{prodcalk} it follows that the solution to
\be \dt \psi = f \psi\ee
is given by
\be \label{sol_calk}
\psi(x)=\exp\left(\int_{\t^{\infty(x)}}^x\frac{-\ln(1-(t-\t(t))f(t))}{t-\t(t)}d_\t t\right) \psi(\t^{\infty}(x)).\ee

Now, we are going to investigate the functional equation of the form

\begin{equation} \label{r1}\left(\begin{array}{@{}c@{}} \dt\psi(x) \\ \dt\phi(x) \end{array}\right)=\tilde\Lam(x)\left(\begin{array}{@{}c@{}}  \psi(x) \\ \phi(x) \end{array}\right)=\left(\begin{array}{@{}cc@{}}  \tilde a(x) & \tilde b(x)\\ \tilde c(x) & \tilde d(x) \end{array}\right)\left(\begin{array}{@{}c@{}}  \psi(x) \\ \phi(x)\end{array}\right).\end{equation}
By the substitution
\begin{equation}\label{lamtylda}\Lam(x)=\mathbb I-(x-\t(x))\tilde\Lam(x)\end{equation}
it can be equivalently written as
\begin{equation} \label{r2} \left(\begin{array}{@{}c@{}} T\psi(x) \\ T\phi(x) \end{array}\right)=\Lam(x)\left(\begin{array}{@{}c@{}} \psi(x) \\ \phi(x) \end{array}\right)=\left(\begin{array}{@{}cc@{}} a(x) & b(x)\\ c(x) & d(x) \end{array}\right)\left(\begin{array}{@{}c@{}} \psi(x) \\ \phi(x)\end{array}\right).\end{equation}
If we introduce the function
\begin{equation} \label{u}u(x):=\frac{\phi(x)}{\psi(x)}, \end{equation}
the equation \eqref{r2} assumes the form
\begin{equation} \label{rhom}u(\t(x))=\frac{d(x)u(x)+c(x)}{b(x)u(x)+a(x)}, \end{equation}
which is equivalent to
\begin{equation} \label{riccati}\dt u(x)= \tilde c(x) + \tilde d(x) u(x) - \tilde a(x) u(\t(x)) - \tilde b(x) u(x)u(\t(x)). \end{equation}
Equation \eqref{riccati} is generalization of  $q$-Riccati equation and thus we will call both \eqref{rhom} and
\eqref{riccati} the \mbox{\bf$\t$-Riccati equation}.

Equation \eqref{eq_func} considered in Section \ref{factor} can be obtained
from \eqref{r2} by putting
\begin{equation}\Lam(x)=\left(\begin{array}{@{}cc@{}} \frac{\lambda-\beta(x)}{\alpha(x)} & -\frac{\gamma(x)}{\alpha(x)}\\ 1 & 0
\end{array}\right).\end{equation}

The formal solution to \eqref{r2} is given in terms of the infinite matrix product
\begin{equation} \label{res}\Lam_\infty(x):=\lim_{n\to\infty}\Lam(\t^n(x))\cdots \Lam(\t^2(x))\Lam(\t(x))\Lam(x)\end{equation}
by the formula
\begin{equation} \label{sol}\left(\begin{array}{@{}c@{}} \psi(x) \\ \phi(x) \end{array}\right)=\Lam(x)_\infty^{-1}\left(\begin{array}{@{}c@{}} \psi(\t^\infty(x)) \\ \phi(\t^\infty(x)) \end{array}\right).\end{equation}
Sometimes we are able to compute resolvent function \eqref{res} explicitly.

\begin{prop}\label{triangle}$\;$
\\
If $a$, $b$ and $d$ are continuous at $\t^\infty(x)$ and $c(x)\equiv 0$ then
\begin{equation} \label{res_triangle}\Lam_\infty(x)=\left(\begin{array}{@{}cc@{}} \exp \int^x_{\t^\infty(x)}\frac{\ln a(t)}{t-\t(t)}d_\t t & F(x) \\
                                            0 & \exp \int^x_{\t^\infty(x)}\frac{\ln d(t)}{t-\t(t)}d_\t t \end{array}\right),\end{equation}
where $F(x)$ is given by
\begin{equation} \label{F}F(x)=\exp \left(\int^x_{\t^\infty(x)}\!\!\frac{\ln d(t)}{t-\t(t)}d_\t t \right)\int^x_{\t^\infty(x)}\!\!\frac{b(t)}{(t-\t(t))a(t)}
\exp \left(\int^t_{\t^\infty(x)}\!\!\frac{\ln \frac{a(s)}{d(s)}}{s-\t(s)}d_\t s \right)d_\t t \end{equation}
\end{prop}

\prf
From the fact that $\Lam$ is upper-triangular we conclude that $\Lam_\infty$ is given by
\begin{equation} \Lam_\infty(x)=\left(\begin{array}{@{}cc@{}} a_\infty(x) & F(x) \\ 0 &  d_\infty(x) \end{array}\right),\end{equation}
where $a_\infty(x)=\prod_{n=0}^\infty T^n a(x)$ and $d_\infty(x)=\prod_{n=0}^\infty T^n d(x)$.
Since
\begin{equation} \Lam_\infty(x)=\Lam_\infty(\t(x))\Lam(x) \end{equation}
we see that $F(x)$ satisfies the equation
\begin{equation} \label{f}F(x)=d(x) F(\t(x)) + b(x) d_\infty(\t(x)). \end{equation}
Solving \eqref{f} we get
\begin{equation} F(x)=\left(\prod_{l=0}^\infty d(\t^l(x))\right)\sum_{k=0}^\infty \frac{b(\t^k(x))}{a(\t^k(x))}\prod_{j=k}^\infty \frac{a(\t^j(x))}{d(\t^j(x))}.\end{equation}
From \eqref{prodcalk}, \eqref{sol_calk} and \eqref{calkdef} we obtain now \eqref{F}. \qed

We should note that if one considers equations \eqref{rhom} or \eqref{riccati} then the matrix-valued function $\Lam$ is
defined up to a function factor and this observation can be used to regularize the product \eqref{res}.

We are going to consider the following gauge-like transformation
\begin{equation} \label{darboux} \Lam(x) \tto \Lam'(x)=D(\t(x))^{-1}\Lam(x)D(x), \end{equation}
for some $D:X\to GL(2,\C)$, which transforms the functional equation \eqref{r2} into the one of the same type with matrix $\Lam'$. Its solutions
are given by
\begin{equation} \label{darboux2}\left(\begin{array}{@{}c@{}} \psi(x) \\ \phi(x) \end{array}\right) \tto D(x)^{-1} \left(\begin{array}{@{}c@{}} \psi(x) \\ \phi(x)
\end{array}\right).\end{equation}
The resolvent of new equation is given by
\begin{equation} \label{dar_res}\Lam_\infty(x) \tto D(0)^{-1}\Lam_\infty(x) D(x). \end{equation}
We call this transform the {\bf$\t$-Darboux transform}. It is a generalization of $q$-Darboux transform introduced in
\cite{ala}. Sometimes we are able to reduce $\Lam$ to such $\Lam'$ that we can compute \eqref{res} explicitly.
This is the case, for example, if $\Lam'$ is upper-triangular. Applying this method, we can obtain the result for
$\t$-Riccati case, which generalize this of $q$-Riccati equation of \cite{ala}.

This is a classical result for the Ricatti equation generalized to $q$-case in \cite{ala} and here it is presented in $\t$-version.
\begin{prop}\label{particular}$\;$ \\
Let $u_0$ be a particular solution of \eqref{rhom}. The general solution of \eqref{r2} is then given by the formula
$$\label{rpsi}\psi(x)=\exp \left(-\!\!\int^x_{\t^\infty(x)}\!\!\frac{\ln a(t)+b(t)u_0(t)}{t-\t(t)}d_\t t\right)\
\left(-B \int^x_{\t^\infty(x)}\frac{b(t)}{a(t)+b(t)u_0(t)} \times\right.$$
\be\times \left.\exp \left(\int^t_{\t^\infty(x)}\!\!\frac{\ln \frac{a(s)+b(s)u_0(s)}{-b(s)u_0(\t(s))+d(s)}}{s-\t(s)}d_\t s\right)\; d_\t
t+A\right)\ee
\begin{equation} \label{rphi}\phi(x)=u_0(x)\psi(x)+B\exp \left(-\!\!\int^x_{\t^\infty(x)}\!\!\frac{\ln -b(t)u_0(\t(t))+d(t)}{t-\t(t)}d_\t t\right), \end{equation}
where the constants $A$ and $B$ are related to the initial conditions in the following way
\begin{equation} A=\psi(\t^\infty(x)) \end{equation}
and
\begin{equation} B=-\psi(\t^\infty(x))u_0(\t^\infty(x))+\phi(\t^\infty(x)). \end{equation}
Solution of \eqref{rhom} is then given by
\begin{equation} \label{ut} u^t(x)=u_0(x)+\frac{t\exp \left(\int^x_{\t^\infty(x)}\!\!\frac{\ln \frac{a(z)+b(z)u_0(z)}{-b(z)u_0(\t(z))+d(z)}}{z-\t(z)}d_\t z\right)}
{1-t
\int^x_{\t^\infty(x)}\!\!\frac{b(z)}{(z-\t(z))(a(z)+b(z)u_0(z))}
\exp \left(\int^z_{\t^\infty(x)}\!\!\frac{\ln \frac{a(s)+b(s)u_0(s)}{-b(s)u_0(\t(s))+d(s)}}{s-\t(s)}d_\t s \right)\;d_\t z
 }\end{equation}
for any $t\in\R$.
\end{prop}
\prf In order to obtain the general solution of \eqref{r1}, we apply the $\t$-Darboux transform \eqref{darboux} with
$D(x)$ of the form
\begin{equation} D(x)=\left(\begin{array}{@{}cc@{}}1 & 0 \\ u_0(x) & 1 \end{array}\right).\end{equation}
Since $u_0$ is a particular solution of \eqref{rhom}, the transformed matrix $\Lam'(x)$ is upper-triangular. Solution
given by \eqref{rpsi} and \eqref{rphi} is obtained by substituting into \eqref{sol} the formulae \eqref{res_triangle}
and \eqref{F} from Proposition \ref{triangle} and transforming formula \eqref{dar_res} for resolvent. Formula \eqref{ut}
follows from \eqref{u}, \eqref{rpsi} and \eqref{rphi} by putting $t:=\frac{B}{A}$. \qed

We define the transform $\B^+_t$ by
\begin{equation} \label{B+}\B^+_t u_0:=u^t. \end{equation}
It is easy to check that $\B^+_t$, $t\in\R$, form a one-parameter group
\begin{equation} \label{Bgroup}B_{s}^+\circ B_{t}^+=B^+_{s+t}. \end{equation}
Moreover the solutions $u^{t_1}(x)$, $u^{t_2}(x)$, $u^{t_3}(x)$ and $u^{t_4}(x)$ do satisfy the unharmonical
superposition principle:
\begin{equation} \label{unharm}\frac{(u^{t_4}(x)-u^{t_3}(x))(u^{t_1}(x)-u^{t_2}(x))}{(u^{t_3}(x)-u^{t_1}(x))(u^{t_2}(x)-u^{t_4}(x))}=
\frac{(t_4-t_3)(t_1-t_2)}{(t_3-t_1)(t_2-t_4)}. \end{equation} The proof of \eqref{Bgroup} and \eqref{unharm} is
straightforward.

Let us mention another application of $\t$-Darboux transform.
When $\Lam(\t^\infty(x))=\mathbb I$ we call \eqref{r2} the regular equation. Solutions of regular equation are finite
and nonzero at $\t^\infty(x)$. Let us introduce singularity (or zero) at this point by the $\t$-Darboux transform
\eqref{darboux} - \eqref{darboux2} with
\begin{equation} \label{dar_sing}D(x)=\left(\begin{array}{@{}cc@{}} (x-\t^\infty(x))^{\delta_1} & 0 \\ 0 & (x-\t^\infty(x))^{\delta_2} \end{array}\right). \end{equation}
Thus one has
\begin{equation} \label{sing}\Lam'(x)=\left(\begin{array}{@{}cc@{}} a(x)\left(\frac{x-\t^\infty(x)}{\t(x)-\t^\infty(x)}\right)^{\delta_1}
 & b(x)\frac{(x-\t^\infty(x))^{\delta_2}}{(\t(x)-\t^\infty(x))^{\delta_1}} \\ c(x)\frac{(x-\t^\infty(x))^{\delta_1}}{(\t(x)-\t^\infty(x))^{\delta_2}} &
 d(x)\left(\frac{x-\t^\infty(x)}{\t(x)-\t^\infty(x)}\right)^{\delta_2}\end{array}\right) \end{equation}
and the solution to \eqref{rhom} with $\Lam'(x)$ is
\begin{equation} \label{meroF}u'(x)=(x-\t^\infty(x))^{\delta_1-\delta_2} u(x),\end{equation}
where $u(x)$ is a solution to \eqref{rhom} with $\Lam(x)$.

Hence one can have $u'(\t^\infty(x))=\infty$ for $\delta_1-\delta_2<0$ while $u(\t^\infty(x))<\infty$.
Thus we can apply the
transformation \eqref{dar_sing} to regularize the singular equation of the type \eqref{sing} to the regular one.
It was applied implicitly for example in \eqref{alpha_sing}.


Infinite products in previous formulae were always assumed to be convergent. We will give now
a condition for this to be the case.
\begin{prop}\label{converg}$\;$\\
Let the following conditions be satisfied:
\begin{enumerate}[i)]
\item $\sum_{k=0}^\infty|\t^k(x)-\t^{k+1}(x)|<\infty $, e.g. it has a place if $\t$ is a contracting
map;
\item $\tilde\Lam$ is continuous at $\t^\infty(x)$.
\end{enumerate}
Then product \eqref{res} is convergent.
\end{prop}
\prf
We will show that under these assumptions $P_n=\Lam(\t^n(x))\cdots\Lam(x)$ is a Cauchy sequence
and thus convergent. We can estimate the norm of the difference between two terms of this series by
\begin{equation} \norm{P_n-P_{n+k}}\leq \norm{\phantom{\tilde\Lam}\Lam(\t^{n+k}(x))\cdots\Lam(\t^{n+1}(x))-1}\norm{P_n}.\end{equation}
Let us show that it can be made arbitrarily small. To that end we show first that $\norm{P_n}$ is
bounded.
\begin{equation} \norm{P_n}\leq\prod_{k=0}^n(1+|\t^k(x)-\t^{k+1}(x)|\norm{\tilde\Lam(\t^k(x))}). \end{equation}
This product is convergent if and only if
\begin{equation} \label{scalarny}\sum_{n=0}^\infty|\t^n(x)-\t^{n+1}(x)|\norm{\tilde\Lam(\t^n(x))}<\infty.\end{equation}
Assumptions of the proposition guarantees that this is the case. Hence
\begin{equation} \norm{P_n}\leq\prod_{k=0}^\infty(1+|\t^k(x)-\t^{k+1}(x)|\norm{\tilde\Lam(\t^k(x))})<\infty.\end{equation}
Now we show that for $n$ big enough  $\norm{\Lam(\t^{n+k}(x))\cdots\Lam(\t^{n+1}(x))-1}$ is arbitrarily
small. We have the inequality
\begin{equation}\label{sccc}\norm{\Lam(\t^{n+k}(x))\cdots\Lam(\t^{n+1}(x))-1}+1\leq \end{equation}
$$\leq\left(1+|\t^{n+k}(x)-\t^{n+k+1}(x)|\norm{\tilde\Lam(\t^{n+k}(x))}\right)\!\cdots\!
\left(1+|\t^{n+1}(x)-\t^{n+2}(x)|\norm{\tilde\Lam(\t^{n+1}(x))}\right). $$ Since \eqref{scalarny} hold then
\eqref{sccc} tends to $1$ as $n\to\infty$. Thus we have completed the proof that $P_n$ is Cauchy sequence. \qed

\section*{Acknowledgement}
Authors would like to thank dr. M. Klimek for the information about works \cite{arai,klimek} and
dr. S. Stepin, dr. M. Horowski for careful reading of the manuscript and interest in the paper. We
would also like to thank Alina Dobrogowska for help and fruitful discussions.
This work is supported in part by KBN grant 2 PO3 A 012 19.

\end{document}